\documentclass[12pt,titlepage]{article}
\usepackage{bm, amssymb,pifont,cancel, amsmath,comment,color}
\usepackage[dvips]{graphicx}
\usepackage[dvipdfm,
colorlinks=true,
linkcolor=blue,
citecolor=blue,
filecolor=blue,
urlcolor=blue
]{hyperref}  

\makeatletter

\def\slash#1{\not\!\!#1}

\@addtoreset{equation}{section}
\makeatother
 
\begin{document}

\begin{titlepage}
\null
\begin{flushright}
WU-HEP-16-21
\end{flushright}

\vskip 1.5cm
\begin{center}
\baselineskip 0.8cm
{\LARGE \bf
More on DBI action in 4D ${\cal N}=1$ supergravity }

\lineskip .75em
\vskip 1.5cm

\normalsize

{\large Shuntaro Aoki} $^{1}${\def\thefootnote{\fnsymbol{footnote}}\footnote[1]{E-mail address: shun-soccer@akane.waseda.jp}}, 
{\large and} {\large Yusuke Yamada} $^{2}${\def\thefootnote{\fnsymbol{footnote}}\footnote[2]{E-mail address: yusukeyy@stanford.edu}}

\vskip 1.0em

{\small\it Department of Physics, Waseda University, \\ 
Tokyo 169-8555, Japan}
\vskip 1.0em

$^2${\small{\it Stanford Institute for Theoretical Physics and Department of Physics,\\
Stanford University, Stanford, CA 94305, U.S.A.}}
\vskip 1.0em

\vspace{12mm}

{\bf Abstract}\\[5mm]
{\parbox{13cm}{\hspace{5mm} \small
We construct a Dirac-Born-Infeld (DBI) action coupled to a two-form field in four dimensional $\mathcal{N}=1$ supergravity. Our superconformal formulation of the action shows a universal way to construct it in various Poincar\'e supergravity formulations. We generalize the DBI action to that coupled to matter sector.  We also discuss duality transformations of the DBI action, which are useful for phenomenological and cosmological applications. 
}}

\end{center}

\end{titlepage}

\tableofcontents
\vspace{35pt}
\hrule
\section{Introduction}\label{intro}
The Dirac-Born-Infeld (DBI) type actions~\cite{Born:1934gh,Dirac:1962iy}, which are nonlinear generalizations of Maxwell theory, naturally appear as the low-energy effective description of D-branes in superstring theory. In particular, if supersymmetry (SUSY) is preserved at low energy scale, D-branes should also be described in a SUSY way. Such attempts have been studied so far (see ,e.g., \cite{Howe:1996mx,Aganagic:1996pe,Bergshoeff:1996tu,Aganagic:1996nn,Bergshoeff:2013pia}). Also, D-brane actions coupled to supergravity (SUGRA) background are discussed in Refs.~\cite{Cederwall:1996pv,Cederwall:1996ri}. These constructions clarify the direct relation between superstring and its effective field theory. 

As a bottom-up approach to superstring theory, 4D ${\cal N}=1$ SUSY/SUGRA model buildings are quite useful, which directly relate the models to collider experiments and cosmological observations. In particular, the current progress of cosmological observation gives constraints on inflation models in the early universe. From this perspective, coupling to not the SUGRA background but the dynamical SUGRA multiplet is important for model buildings. 

In this work, we develop further the DBI type actions in 4D ${\cal N}=1$ SUGRA, which have been discussed in Refs.~\cite{Cecotti:1986gb,Kuzenko:2002vk,Kuzenko:2005wh,Kuzenko:2009ym,Abe:2015nxa,Abe:2015fha}. The effective action of a single D3-brane would take the following form,
\begin{align}
 \sim\int d^4x \sqrt{-{\rm det}(g_{\mu\nu}+\partial_\mu\phi^i\partial_\nu\phi^{*i}+F_{\mu\nu}+B_{\mu\nu})},
 \end{align}
where $g_{\mu\nu}$ is a graviton, $\phi^i$ are position moduli, $F_{\mu\nu}$ is a field strength of an abelian gauge field, and $B_{\mu\nu}$ is the so-called B-field, an antisymmetric tensor. Although some parts of this action have been realized within the off-shell ${\cal N}=1$ SUGRA formulation, there is no complete realization of it. 

What we focus on in this paper is the couplings to B-field in SUGRA, which has never been discussed in previous works. The B-field, which comes from the NS-NS sector in superstring, is necessary for, e.g., Green-Schwartz anomaly cancellation~\cite{Green:1984sg}. For the SUGRA realization of DBI action, the observation in partial breaking of ${\cal N}=2 $ SUSY is very helpful, which is discussed, e.g., in Refs.~\cite{Hughes:1986dn,Hughes:1986fa,Bagger:1996wp,Rocek:1997hi}. There are various discussions related to the DBI actions in Refs.~\cite{Ketov:1998ku,Kuzenko:2000uh,Ferrara:2014oka,Ferrara:2014nwa,Andrianopoli:2014mia,Andrianopoli:2015wqa,Andrianopoli:2015rpa,Andrianopoli:2016eub}. In Refs.~\cite{Kuzenko:2000uh,Ambrosetti:2009za,Ferrara:2015ixa}, the combination including $F+B$ was constructed in global SUSY, and also discussed some duality properties the action has. We first embed it into superconformal system, which unifies the construction of an action in various SUGRA formulations~\cite{Kaku:1978nz,Kaku:1978ea,Townsend:1979ki,Kugo:1982cu,Kugo:1983mv}. Then, based on the minimal formulation, we generalize it to the matter coupled system. 

We also show the bosonic component expressions of the action for applications particularly to cosmology. It has been known that the SUSY higher-order derivative terms, which the DBI action also has, show interesting features and play important roles in cosmological model buildings. For example, such terms of chiral superfields have been investigated in Refs.~\cite{Khoury:2010gb,Khoury:2011da,Baumann:2011nm,Koehn:2012ar,Farakos:2012je,Farakos:2012qu,Koehn:2012te,Sasaki:2012ka,Farakos:2013zya,Gwyn:2014wna,Dalianis:2014nwa,Dalianis:2014sqa,Aoki:2014pna,Aoki:2015eba,Ciupke:2015msa,Bielleman:2016grv,Ciupke:2016agp,Kimura:2016irk}. We also show the duality relations, which enable us to see the system from different viewpoints. As we will see, the DBI action with the B-field is dual to the massive DBI action, which has been shown in Refs.~\cite{Abe:2015fha,Ambrosetti:2009za,Ferrara:2015ixa}. 

The remaining parts of this paper are organized as follows. In Sec.~\ref{global}, we first review the construction of DBI action with and without couplings to the B-field. We will see that the constraint imposed between two chiral superfields and the structure of nilpotency are important for the constructions. Then, we discuss the SUGRA generalization of the DBI with the B-field in Sec.~\ref{extension}. We briefly review the basics of conformal SUGRA in Sec.~\ref{preparation}. Then we embed the constraint, which is a key to realize the DBI action, into conformal SUGRA, and also extend it to a matter coupled system in Sec.~\ref{embedding}. We will find that in all the SUGRA formulations, we can realize the DBI action in a unified manner. Also, its bosonic component expansions will be shown in the old minimal case in Sec.~\ref{component}. The resultant action has some dual properties which we discuss in Sec.~\ref{dual}.  Finally, we summarize this paper in Sec.~\ref{summary}. In Appendix.~\ref{AUX}, we show the details for deriving the on-shell Lagrangian. In Appendix.~\ref{DUAL}, the explicit component expressions of dual actions obtained in Sec.~\ref{dual} are shown.

Throughout this paper, we use Planck unit $M_P=1$ where $M_P=2.4\times 10^{18}$ GeV is the reduced Planck mass.
\section{DBI action in global SUSY}\label{global}
Here, we briefly review the DBI action with and without couplings to a two-form superfield containing the B-field in global SUSY, which we will embed into conformal SUGRA formulation in Sec.~\ref{extension}. Since the superfield formulation is rather simpler than the $\rm{Poincar\acute{e}}$ SUGRA case discussed later, it would be useful to understand the construction in global SUSY case.

\subsection{DBI action without B-field}\label{global model}
On the basis of Refs.~\cite{Bagger:1996wp,Rocek:1997hi}, we first review the construction of SUSY DBI action including the field strength of a U(1) vector potential $A_a$,  
\begin{align}
 L&=1-\sqrt{-{\rm{det}}\left( \eta_{ab}+  F_{ab} \right)}, \label{F}
\end{align}  
where $F_{ab}=\partial _a A_b-\partial _b A_a$.
To embed this Lagrangian into superfield formalism\footnote{We follow the superspace conventions of Ref.~\cite{Wess:1992cp}.}, we need a real superfield $V$ and its field strength superfield $W_{\alpha }\equiv -\frac{1}{4}\bar{D}^2D_{\alpha }V$ where
\begin{align}
W_{\alpha }=-i\lambda _{\alpha }+\left( \delta _{\alpha }^{\beta }D-\frac{i}{2}(\sigma ^{a}\bar{\sigma }^{b})_{\alpha } ^{\ \beta }F_{ab}\right) \theta _{\beta }+\theta \theta \sigma _{\alpha \dot{\alpha }}^a\partial _a\bar{\lambda }^{\dot{\alpha }}. \label{CFS}
\end{align}
$D_{\alpha }$ and $\bar{D}_{\dot{\alpha }}$ are a spinor derivative and its conjugate. $\lambda $ is a Weyl spinor and $D$ is a real scalar. Here we denote $D^2=D^{\alpha }D_{\alpha }$, $\bar{D}^2=\bar{D}_{\dot{\alpha }}\bar{D}^{\dot{\alpha }}$.

The key for constructing a SUSY DBI action is the following constraint imposed between chiral superfields $X$ and $W_{\alpha}$, 
\begin{align}
X+aX\bar{D}^2\bar{X}+bW^2=0, \label{constraint 1}
\end{align} 
where $a$ and $b$ are real parameters. Note that this condition observed in Refs.~\cite{Bagger:1996wp,Rocek:1997hi} is a consequence of partially broken ${\cal N}=2$ SUSY. Although the view of the partial SUSY breaking gives physically interesting understanding, we just use this constraint as a tool to realize the DBI action. 

The constraint can be solved with respect to $X$ as a nonlinear function of $W^2$,
\begin{align}
X(W,\bar{W})= -bW^2-2ab^2\bar{D}^2\biggl[ \frac{W^2\bar{W}^2}{1-2abA\pm \sqrt{1-4abA+4a^2b^2B^2}}\biggr] , \label{sol F}
\end{align}
where 
\begin{align}
A\equiv \frac{1}{2}(D^2W^2+\bar{D}^2\bar{W}^2),\ \ \ \ B\equiv \frac{1}{2}(D^2W^2-\bar{D}^2\bar{W}^2).
\end{align}
This is a consequence of the nilpotency of the third term $W^2$ in Eq.$\eqref{constraint 1}$. Also, the solution $X(W,\bar{W})$ is proportional to $W^2$, which leads to $X^2=0$. This shows the underlying Volkov-Akulov nonlinear SUSY~\cite{Volkov:1972jx,Volkov:1973ix,Rocek:1978nb,Lindstrom:1979kq}.

Using the solution $\eqref{sol F}$, one can obtain the DBI action from the linear term of $X(W,\bar{W})$,
\begin{align}
L=\int d^2\theta X (W,\bar{W})+{\rm{h.c.}} . \label{global F}
\end{align} 
After integrating out the auxiliary field $D$, the bosonic part of Eq. $\eqref{global F}$ gives 
\begin{align}
\nonumber L|_{boson}=&\frac{1}{4a}\Biggl[ 1- \sqrt{1-8abF_{ab}F^{ab}+16a^2b^2(F_{ab}\tilde{F}^{ab})^2}\Biggr] \\
=&\frac{1}{4a}\Biggl[ 1- \sqrt{-{\rm{det}}\left( \eta_{ab}+4\sqrt{-ab}  F_{ab} \right)}\Biggr] ,\label{global FF}
\end{align}
where $\tilde{F}_{ab}\equiv -\frac{i}{2}\varepsilon _{abcd}F^{cd}$. Here, we have used the following formula
\begin{align}
-{\rm{det}}\left(   \eta_{ab}+\alpha   F_{ab} \right) =1+\frac{\alpha ^2}{2}F_{ab}F^{ab}+\frac{\alpha ^4}{16}(F_{ab}\tilde{F}^{ab})^2,
\end{align}
and chosen the ``$+$" branch in Eq. $\eqref{sol F}$.
For $a=1/4$ and $b=-1/4$ in Eq. $\eqref{global FF}$, we obtain
\begin{align}
L|_{boson}= 1-\sqrt{-{\rm{det}}\left( \eta_{ab}+ F_{ab} \right)} ,
\end{align}
which is exactly the DBI form $\eqref{F}$.

To extend the action to ,e.g., matter coupled system, it is rather useful to impose the constraint $\eqref{constraint 1}$ by a Lagrange multiplier chiral superfield $\Lambda $ as~\cite{Rocek:1997hi}, 
\begin{align}
L=\int d^2 \theta \Biggl[ X+ \Lambda \left( X+aX\bar{D}^2\bar{X}+bW^2 \right) +MX^2\Biggr] +{\rm{h.c.}}. \label{global F2}
\end{align}
Here we have also introduced the last term with a multiplier $M$, whose variation gives $X^2=0$ trivially satisfied by a solution $\eqref{constraint 1}$. Due to $X^2=0$, the lowest component of $X$ becomes a fermion bilinear which simplifies our computations, as long as we extract bosonic parts. This additional constraint neither conflicts with Eq. $\eqref{constraint 1}$ nor imposes overconstraint on this system. 
 
 
\subsection{DBI action with B-field} \label{2form}
Now, let us move to the case where a two-form field $B_{ab}$ is coupled to a U(1) vector. Such a two-form is naturally described by a real linear superfield $L$ which satisfies
\begin{align}
D^2L=\bar{D}^2L=0. \label{def linear}
\end{align}
Its component expression is 
\begin{align}
L=C+\theta \eta +\bar{\theta }\bar{\eta }+\theta \sigma ^a\bar{\theta }B_a-\frac{i}{2}(\theta \theta )\bar{\theta }\bar{\sigma }^a\partial _a\eta -\frac{i}{2}(\bar{\theta }\bar{\theta })\theta \sigma ^a\partial _a \bar{\eta }-\frac{1}{4}\theta \theta \bar{\theta }\bar{\theta }\Box C ,
\end{align}
where $C$ is a real scalar, $\eta $ is a Weyl fermion, and a real vector $B_a$ satisfies $\partial ^aB_a=0$, that is, it can be written using a two-form field $B_{ab}$ as $B_a= \frac{1}{2}\varepsilon _{abcd}\partial ^bB^{cd}$. The definition $\eqref{def linear}$ implies the existence of the chiral spinor prepotential $\Phi _{\alpha }$~\cite{Siegel:1979ai},
\begin{align}
L=D^{\alpha }\Phi _{\alpha }+\bar{D}_{\dot{\alpha} }\bar{\Phi} ^{\dot{\alpha} },\ \ \ \ \bar{D}_{\dot{\beta }}\Phi _{\alpha }=0, \label{defphi}
\end{align}
with component expansion,
\begin{align}
\Phi _{\alpha }=\xi  _{\alpha }-\theta _{\beta }\left( \frac{1}{2}\delta _{\alpha }^{\beta }(C+iE)+\frac{1}{4}(\sigma ^a \bar{\sigma }^b)_{\alpha }^{\ \beta } B_{ab}\right) +\theta \theta \left(  \eta _{\alpha }+i\sigma ^a_{\alpha \dot{\alpha }}\partial_a\bar{\xi }^{\dot{\alpha }}\right) ,
\end{align}
where $\xi  $ is a Weyl fermion and $E$ is a real scalar. Note that, due to the identity $D^{\alpha }\bar{D}^2D_{\alpha }-\bar{D}_{\dot{\alpha }}D^2\bar{D}^{\dot{\alpha }}=0$, $L$ is invariant under the gauge transformation,
\begin{align}
\Phi _{\alpha }\rightarrow \Phi _{\alpha }+\frac{i}{8}\bar{D}^2D_{\alpha }\Theta ,  \label{GT2}
\end{align}
where $\Theta $ is a real superfield. By this gauge transformation, $E$ and $\xi $ can be gauged away.

In Sec.~\ref{global model}, we see that the constraint $\eqref{constraint 1}$ with the nilpotency of $W^2$ is important for the construction. By this observation, we expect a combination 
\begin{align}
(W-2ig\Phi )^2,
\end{align} 
could be replaced with $W^2$, since it still possesses the required nilpotency. Here, $g$ is a real constant. Furthermore, this combination is invariant under the gauge transformation $\eqref{GT2}$ if we assign the following transformation,
\begin{align}
V\rightarrow V+g\Theta ,\ \ \ \  W_{\alpha } \rightarrow W_{\alpha }-\frac{g}{4}\bar{D}^2D_{\alpha }\Theta .
\end{align}
 to the vector superfield $V$. 
The DBI Lagrangian coupled with the two-form can be written as a natural generalization of Eq. $\eqref{global F2}$, 
\begin{align}
L=\int d^2 \theta \Biggl[ X+ \Lambda \left( X+aX\bar{D}^2\bar{X}+b(W-2ig\Phi )^2 \right) +MX^2\Biggr] +{\rm{h.c.}} +L_{kin,L},  \label{global B+F}
\end{align}
where $L_{kin,L}$ is responsible for the kinetic term of $L$, though we do not specify it here. 

Finally, let us comment on the relation to the partial ${\cal N}=2$ SUSY breaking. This action can also be realized within partially broken ${\cal N}=2$ SUSY, where a Goldstino $\mathcal{N}=2$ vector superfield couples to a ${\cal N}=2$ tensor multiplet~\cite{Ambrosetti:2009za}. Then, applying dual transformation to supersymmetric $B\wedge F$ coupling~\cite{Ferrara:2015ixa}, we obtain the same action. Interestingly, the Goldstino superfield $W_\alpha$ can be eaten by the two-form superfield by super gauge transformation~$\eqref{GT2}$. This means the super-Higgs mechanism works without a gravitino in ${\cal N}=2$ SUGRA~\cite{Ambrosetti:2009za}. 
\section{Extension to 4D $\mathcal{N}=1$ conformal SUGRA}\label{extension}
In this section, we discuss the SUGRA generalization of Eq.~$\eqref{global B+F}$ in conformal SUGRA, which is briefly reviewed in Sec.~\ref{preparation}. Using this technique, we embed the constraint and realize (matter coupled) action in Sec.~\ref{embedding}. We will see that our action can be realized in any SUGRA formulations. Then, we derive component expressions especially in the old minimal formulation as an example.
We follow the conventions of~\cite{Freedman:2012zz} here and hereafter.  

\subsection{Preliminary}\label{preparation}
Here, we briefly review conformal SUGRA and show multiplets necessary for our purpose (see Refs.~\cite{Kugo:1982cu,Kugo:1983mv,Freedman:2012zz} for more details).

Conformal SUGRA is a gauge theory of superconformal group which consists of dilatation, $U(1)_A$ symmetry, S-SUSY and conformal boost in addition to symmetries of Poincar\'e SUGRA (translation, SUSY and Lorentz symmetry). First we construct an action invariant under superconformal symmetry, and break symmetries spontaneously other than $\rm{Poincar\acute{e}}$ SUSY by imposing gauge fixing conditions. Some of them are imposed on an unphysical multiplet called a compensator multiplet. This conformal SUGRA technique enables us to circumvent complicated field redefinitions which are present in Poincar\'e SUGRA methods, thanks to extra symmetries in superconformal group. Furthermore, it gives us a unified description of different SUGRA formulations such as the old and new minimal formulations, which have different sets of auxiliary fields in the gravity multiplet. In terms of conformal SUGRA, these differences come from the choices of compensator superfields, e.g., the chiral compensator $S_0$ leads to the old minimal formulation, and a real linear compensator $L_0$ does the new minimal formulation\footnote{We distinguish the physical multiplet and compensator by adding a subscript $"0"$ to the latter.}.  

One more important notion is that the multiplet in conformal SUGRA is characterized by charges under the dilatation and $U(1)_A$ symmetry, called Weyl weight $w$ and chiral weight $n$, respectively. Typical multiplets have specific Weyl and chiral weights. For example, a chiral multiplet $X$ which is defined by
\begin{align}
\bar{\mathcal{D}}_{\dot{\alpha}}X=0, 
\end{align}
satisfies a weight condition, $w=n$. Here, $\bar{\mathcal{D}}_{\dot{\alpha}}$ is a spinor derivative corresponding to the superconformal generalization of $\bar{D}_{\dot{\alpha}}$ \cite{Kugo:1983mv}. Also, weights of anti-chiral multiplet $\bar{X}$ must satisfy $w=-n$. On the other hand, a real linear multiplet $L$ which satisfies 
\begin{align}
\Sigma L =\bar{\Sigma } L=0, \label{defL}
\end{align}
has determined weights as $(w,n)=(2,0)$. $\Sigma $ and $\bar{\Sigma }$ are chiral projection operator which is analogous operation of $\bar{D}^2$ and $D^2$ in global SUSY \cite{Kugo:1983mv}. $\Sigma $ ($\bar{\Sigma }$) can be applied when the operand satisfies the weight condition $w-n=2$ ($w+n=2$), which is consistent with the weights of $L$.

These multiplets, $X$ and $L$, can be considered as irreducible representations of general multiplet $\Phi$ with weights $(w,n)$, whose components are given by
\begin{align}
\Phi =\{ \mathcal{C}, \mathcal{Z},\mathcal{H},\mathcal{K},\mathcal{B}_a,\Lambda ,\mathcal{D}\} .\label{general multiplet}
\end{align}
Here $\mathcal{Z}$ and $\Lambda $ are Dirac spinors, $\mathcal{B}_a$ is a vector, and the others ($\mathcal{C},\mathcal{H},\mathcal{K},\mathcal{D}$) are complex scalars. Indeed, the chiral multiplet $X$ is embedded into a general multiplet as 
\begin{align}
X=\{ X,-\sqrt{2}iP_L\chi , F_X,iF_X,iD_aX,0,0  \}, \label{embedc}
\end{align}
 where $X$ and $F_X$ are complex scalars\footnote{We use same letter for the first component of chiral multiplet.}, and $\chi$ is a Majorana spinor. $P_L$ is a left-handed projection operator, thus $P_L\chi$ is Weyl spinor. $D_a$ denotes a superconformal covariant derivative. Instead, the real linear multiplet $L$ is embedded as
\begin{align}
L=\{ C,Z,0,0,B_a,-\slash{D}Z,-\Box C\} , \label{embedl}
\end{align}
where $C$ is a real scalar, $Z$ is a Majorana spinor, and $B_a$ is a vector constrained by $D^aB_a=0$. Here we used the conventions $\slash{D}=\gamma ^aD_a$ and $\Box=D^aD_a$.

In the same way with Eq. $\eqref{defphi}$ in global SUSY, one can define the chiral spinor prepotential $\Phi_{\alpha }$ of $L$. Taking into account the fact that $\mathcal{D}$ and $\bar{\mathcal{D}}$ alter weights $(\frac{1}{2},-\frac{3}{2})$ and $(\frac{1}{2},\frac{3}{2})$ \cite{Kugo:1983mv}, we find $\Phi_{\alpha }$ has weights $(\frac{3}{2},\frac{3}{2})$. This multiplet has the following embedding into the chiral multiplet $\eqref{embedc}$~\cite{Cecotti:1987nw},
\begin{align}
\bar{\Omega } \Phi  =\left\{ \bar{\Omega }P_L\eta  , -\frac{i}{4\sqrt{2}}\left(B_{ab}\gamma ^{ab}+C+iE\right) P_L\Omega ,-\frac{1}{2}\bar{\Omega } P_LZ-\bar{\Omega } P_L\slash{D}\eta \right\} , \label{pre}
\end{align}
where we inserted a constant spinor $\Omega $, and $\bar{\Omega }$ is defined by $\bar{\Omega }=\Omega ^{T}C_4$ with charge conjugation matrix $C_4$. The scalar $E$ and two-form $B_{ab}$ are real, and $\eta $ is a Majorana spinor. 

Finally, the chiral field strength multiplet $\mathcal{W}_{\alpha }$ which is a counterpart of Eq. $\eqref{CFS}$ has following structure,
\begin{align}
\bar{\Omega } \mathcal{W} =\left\{ \bar{\Omega }P_L\lambda ,\frac{1}{\sqrt{2}}\left(  -\frac{1}{2}\gamma _{ab}F^{ab}+iD\right) P_L\Omega  ,\bar{\Omega  } P_L\slash{D}\lambda \right\} ,
\end{align}
where $D$ is a real scalar, $\lambda $ is a Majorana spinor, and $F_{ab}$ is a field strength. $\mathcal{W}_{\alpha }$ also has weights $(\frac{3}{2},\frac{3}{2})$. As is the global SUSY case, the combination $\mathcal{W}_{\alpha }-2ig\Phi_{\alpha }$ is gauge invariant under the transformations
\begin{align}
\delta \mathcal{W}_{\alpha }=-\frac{g}{2}\Sigma \mathcal{ D}_{\alpha }\Theta,\ \ \ \ \delta  \Phi_{\alpha }=\frac{i}{4}\Sigma \mathcal{ D}_{\alpha }\Theta, \label{gaugetheta}
\end{align}
where $\Theta$ is a real multiplet with weights $(0,0)$.

For our purpose, we enumerate some formulas of conformal SUGRA such as action formulas and multiplication rule, mainly focusing on bosonic parts \footnote{Useful formulas in conformal SUGRA are summarized in Ref.~\cite{Ferrara:2016een}.}. For chiral multiplet with weights $(w,n)=(3,3)$, we have the superconformal F-term formula, 
\begin{align}
[X]_F=\int d ^{4}x\sqrt{-g}\frac{1}{2}(F_X+{\rm{h.c.}}) +{\rm{fermion\ terms}}. \label{Fformula}
\end{align}

For a real multiplet with $(w,n)=(2,0)$ whose contents are $\phi =\{C,Z,H,K,B_a,\Lambda,D\}$ with all components real (Majorana), superconformal D-term formula can be applied,
\begin{align}
[\phi ]_D= \int d^{4}x\sqrt{-g}\left(  D -\frac{1}{3}CR \right) +{\rm{fermion\ terms}} ,\label{Dformula}
\end{align}
where $R$ denotes Ricci scalar.
In our notation, each formula is connected by
\begin{align}
[\phi ]_D=[\Sigma (\phi )]_F.
\end{align}

We also need multiplication rule. The multiplet whose first component is given by $f({\cal C}^I) $, where $I$ classifies different multiplets, has following form
\begin{align}
\nonumber \biggl\{ &f,..., f_I\mathcal{H}^I+..., f_I\mathcal{K}^I+..., f_I\mathcal{B}_a^I+...,\\
 &...,f_I\mathcal{D}^ I+\frac{1}{2}f_{IJ}\left( \mathcal{K}^I\mathcal{K}^J+\mathcal{H}^I\mathcal{H}^J-\mathcal{B}^{aI}\mathcal{B}_a^J-D_a\mathcal{C}^ID^a\mathcal{C}^J\right) \biggr\} , \label{formulaF}
\end{align}
where the subscript $I$ means the derivative with respect to $\mathcal{C}^I$, and ellipses denote fermionic terms.

\subsection{Embedding of DBI action into conformal SUGRA}\label{embedding}
We consider embedding of the global constraint,
\begin{align}
X+aX\bar{D}^2\bar{X}+b(W-2ig\Phi )^2=0, \label{constraint 2}
\end{align} 
into conformal SUGRA. Recall that total Weyl and chiral weights of each term in the constraint must be equal. The third term in Eq. $\eqref{constraint 2}$, which would be replaced by $(\mathcal{W}-2ig\Phi )^2$, has weights $(3,3)$ as shown in the previous subsection. Then, weights of the first term $X$ are determined as $(3,3)$. For the second term, however, we cannot replace $\bar{D}^2$ with the chiral projection operator $\Sigma $, since now $\bar{X}$ has weights $(3,-3)$ which conflicts with the weight condition that the operand of $\Sigma $ must satisfy : $w-n=2$. Then, to apply $\Sigma $ correctly, we define
\begin{align}
\bar{X}/\mathcal{U}_0^2, \label{insert}
\end{align}
where $\mathcal{U}_0$ is a general multiplet with weights $(x,x-2)$, thus this multiplet has weights $(w'=3-2x,n'=1-2x)$. Here, $x$ is arbitrary. In this case, Eq. $\eqref{insert}$ satisfies the weight condition $w'-n'=2$, which is consistent with the requirement. Then, we conclude that the second term in Eq. $\eqref{constraint 2}$ should be replaced by $X\Sigma (\bar{X}/\mathcal{U}_0^2)$. 

The corresponding constraint of $\eqref{constraint 2}$ in conformal SUGRA is then,
\begin{align}
X+aX\Sigma (\bar{X}/\mathcal{U}_0^2)+b(\mathcal{W}-2ig\Phi )^2=0. \label{constraint 5}
\end{align}
Since $X\Sigma (\bar{X}/\mathcal{U}_0^2)$ should have $(w,n)=(3,3)$, one find that only a possible value of $x$ is $x=2$. Some comments are as follows : First, we assumed $\mathcal{U}_0$ is a compensator multiplet, since it disappears in the global limit as Eq. $\eqref{constraint 2}$, and this means $\mathcal{U}_0$ should be regarded as gravitational corrections. Secondly, at this stage, we do not specify the formulation of SUGRA, that is, what the compensator $\mathcal{U}_0$ is. Therefore, various SUGRA embeddings can be achieved under different choices of  $\mathcal{U}_0$, as long as their weights have the form as $(2,0)$. For example, $\mathcal{U}_0=S_0\bar{S}_0$, where $S_0$ is a chiral compensator with $(1,1)$, realizes old minimal formulation. On the other hand, $\mathcal{U}_0=L_0$ is also available as a choice of $\mathcal{U}_0$, where $L_0$ is a linear compensator with $(2,0)$, which describes the new minimal formulation. Further, there exists one more off-shell formulation known as non minimal formulation which corresponds to the choice of a complex linear multiplet $\Psi _0$ as a compensator. Here, a complex linear multiplet is defined by relaxing the reality condition of $L$ in Eq. $\eqref{defL}$. $\Psi_0 $ has weights $(w,w-2)$, thus it can be used as $\mathcal{U}_0$, if we also use its conjugate $\bar{\Psi}_0$ with weights $(w, -w+2)$.     

Using Eq. $\eqref{constraint 5}$, the minimal extension of Eq. $\eqref{global B+F}$ is given by
\begin{align}
S=2\biggl[X+\Lambda \left( X+aX\Sigma\left( \frac{ \bar{X}}{\mathcal{U}_0^2} \right) +b(\mathcal{W}-2ig\Phi )^2  \right) + MX^2\biggr]_F +S_{L}, \label{local B+F}
\end{align}
where $\Lambda $ and $M$ are Lagrange multiplier chiral multiplets with weights $(0,0)$ and $(-3,-3)$ respectively. $S_{L}$ is responsible for the kinetic terms of graviton, gravitino and linear multiplet $L$, and takes the following form
\begin{align}
S_{L}=[\mathcal{U}^{\frac{1}{2}}_0\bar{\mathcal{U}}^{\frac{1}{2}}_0\mathcal{F} (\ell)]_D, \ \ \ \ \ell \equiv \frac{L}{\mathcal{U}^{\frac{1}{2}}_0\bar{\mathcal{U}}^{\frac{1}{2}}_0},\label{SL}
\end{align}
where $\mathcal{F}$ is a real function with weights $(0,0)$. 

The minimal action $\eqref{local B+F}$ can be extended to matter coupled system as
\begin{align}
S_M=2\biggl[f(S^i)X+\Lambda \left( X+X\Sigma\left( \frac{ \omega (\ell,S^i,\bar{S}^{\bar{j}}) \bar{X}}{\mathcal{U}_0^2} \right) +b(\mathcal{W}-2ig\Phi )^2  \right) + MX^2\biggr]_F +S_{L,M},\label{local B+F matter}
\end{align}  
where $f$ is a holomorphic function of matter chiral multiplets $S^i$ with $(0,0)$, and $\omega$ is a general function of linear multiplet $\ell$ in addition to matter multiplets. Here, $f$ and $\omega$ have weights $(0,0)$ since $\ell$, $S^i$, and $\bar{S}^{\bar{j}}$ are all $(0,0)$ multiplets. Note that other insertions of matter function, e.g., to the second term $\Lambda X\rightarrow g(S^i) \Lambda X$ do not change the system essentially, since suitable field redefinition leads to Eq. $\eqref{local B+F matter}$ again. Finally, $S_{L,M}$ is a generalization of Eq. $\eqref{SL}$ whose form is 
\begin{align}
S_{L,M}=[\mathcal{U}^{\frac{1}{2}}_0\bar{\mathcal{U}}^{\frac{1}{2}}_0\mathcal{F}(\ell, S^i,\bar{S}^{\bar{j}})]_D. \label{SLM}
\end{align} 
In addition to Eq. $\eqref{SL}$, this part also produces kinetic terms for matter multiplets $S^i$ and $\bar{S}^{\bar{j}}$, governed by K\"ahler potential. 

Furthermore, in the old minimal case, the general form of superpotential $W(S^i)$ can be allowed without any restrictions as
\begin{align}
[S_0^3W(S^i)]_F. \label{W}
\end{align}
Here, $W$ is a  holomorphic function of $S^i$ with $(0,0)$.
In the other formulations such as the new and non minimal, we can also introduce superpotential but restricted class~\cite{Ferrara:1983dh}.

\subsection{Component expansion}\label{component}
In this subsection, we only give a result for the component expressions of our actions, focusing on bosonic fields. 
The minimal action $\eqref{local B+F}$ can be regarded as a special case of a general action $\eqref{local B+F matter}$ with $f\rightarrow 1$, $\omega \rightarrow a$, and matter dependences of Eq. $\eqref{SLM}$ omitted, thus we start from the  derivation for Eq. $\eqref{local B+F matter}$. 

So far, we have not specified the formulation of SUGRA. Here, we consider the old minimal case, taking $\mathcal{U}_0=|S_0|^2$. Following Ref.~\cite{Derendinger:1994gx}, we formally write the functional form of $\mathcal{F}$ as
\begin{align}
\mathcal{F}(\ell , S^i,\bar{S}^{\bar{j}})=\ell G(e^{\frac{K}{3}}\ell ) , \label{ellG}
\end{align}
where $G$ is a real function and $K=K( S^i,\bar{S}^{\bar{j}})$ is a would-be K\"ahler potential. Now, $\ell$ should be understood as $\ell =\frac{L}{|S_0|^2}$. For simplicity, we also assume $\omega $ is real. 
Under these assumptions and simplifications,  we eliminate auxiliary fields and impose superconformal gauge fixing conditions (see appendix~\ref{AUX} for detailed procedure). The bosonic component expressions of Eq. $\eqref{local B+F matter}$ with the superpotential $\eqref{W}$ are given by\footnote{Lagrangian density is defined by $S=\int d^4x\sqrt{-g} \mathcal{L}$.} 
\begin{align}
\nonumber  \mathcal{L}_M=&\frac{1}{2}R -K_{i\bar{j}}\partial _{\mu}S^i \partial ^{\mu}\bar{S}^{\bar{j}} -\frac{3}{4C^2}\frac{2G'+yG''}{G'+yG''}\left( \partial_{\mu}C \partial^{\mu}C -B_{\mu}B^{\mu}\right) -V_F\\
\nonumber  &-\frac{|S_0|^4f_R}{2\omega }+\frac{2ibf_I}{1-\frac{2b\omega }{|S_0|^4}(gC)^2}\mathcal{F}_{ab}\tilde{\mathcal{F}}^{ab} \\
&+\frac{|S_0|^4}{2\omega }\sqrt{f_R^2-\frac{2b\omega }{|S_0|^4}(gC)^2(f_R^2+f_I^2)}\times \sqrt{-{\rm{det}}\left\{ g_{ab}+\sqrt{\frac{-8b\omega /|S_0|^4}{1-\frac{2b\omega }{|S_0|^4}(gC)^2}}\mathcal{F}_{ab}\right\} } , \label{final}
\end{align} 
where 
\begin{align}
V_F=\frac{3|S_0|^4}{2\ell y}\Biggl[\frac{1}{G'}K^{i\bar{j}}(W_i+K_iW)(\bar{W}_{\bar{j}}+K_{\bar{j}}\bar{W})+\frac{3}{G'+yG''}|W|^2\Biggr] . \label{V}
\end{align}
We have used the same letters for the first components of multiplet $f=f_R+if_I$, $\omega $, $\ell$ and compensator $S_0$. Also, subscripts $i$ and $\bar{j}$ denote derivatives with respect to $S^i$ and $\bar{S}^{\bar{j}}$, e.g., $K_{i\bar{j}}=\partial ^2K/\partial S^i\partial \bar{S}^{\bar{j}}$, and $K^{i\bar{j}}\equiv (K_{i\bar{j}})^{-1}$.
$y$ is defined by $y=  e^{\frac{K}{3}}\ell $, and prime denotes a derivative with respect to $y$.
Further, we use gauge invariant combination $\mathcal{F}_{ab}\equiv F_{ab}+gB_{ab}$ and $\tilde{\mathcal{F}}^{ab}\equiv -\frac{i}{2}\varepsilon ^{abcd}\mathcal{F}_{cd}$.  

In Eq. $\eqref{final}$ and $\eqref{V}$, $S_0$ should be understood as the function of physical fields such as $C$ and $S^i$. 
This is because superconformal gauge fixing produces an implicit equation with respect to compensator $S_0$, which one cannot solve in general~\cite{Derendinger:1994gx}. Note that $\mathcal{F}$ has compensator dependence through $\ell =C/|S_0|^2$. Once the functional form of $\mathcal{F}$ ($G$ in this case) is determined, we can fix the value of $S_0$. 

One can easily obtain the component expressions in the minimal case $\eqref{local B+F}$, taking $W=K=0, \omega=a, f=1$. As an simple example, here we consider the case where $\mathcal{F}$ is given by $\mathcal{F}=\ell^{3/2}$~\cite{Derendinger:1994gx}. In this case, the explicit value of $S_0$ is determined from Eq. $\eqref{Dcondition}$ and Eq. $\eqref{Acondition}$ as,
\begin{align}
S_0=\frac{1}{3}C^{\frac{3}{2}} .
\end{align}
In this case, the action $\eqref{final}$ reduces to
\begin{align}
\nonumber  \mathcal{L}=&\frac{1}{2}R -\frac{9}{4C^2}\left( \partial_{\mu}C \partial^{\mu}C -B_{\mu}B^{\mu}\right) \\
&-\frac{M_*^4}{2}C^6\biggl[ 1-\sqrt{1-\frac{g^2}{2M_*^4C^4}}\times \sqrt{-{\rm{det}}\left\{ g_{ab}+\sqrt{\frac{-2 }{M_*^4C^6-\frac{(gC)^2}{2}}}\mathcal{F}_{ab}\right\} } \biggr] ,
\end{align}
where we fixed constants as $3^4a=1/M_*^4$, $b=1/4$. 
\section{Duality relations} \label{dual}
Here, we show some dual descriptions of the action shown in the previous section.  So far, we have derived the possible couplings between an abelian gauge field and the two-form B-field. By the duality relations discussed below, we can see the system from some other perspectives, which would be useful to understand the physical consequences and to compare it with superstring effective actions. 
\subsubsection*{Electric-magnetic duality}
First, we show that the Eq. $\eqref{local B+F matter}$, which is the DBI action of a two-form ${\cal F}_{ab}=F_{ab}+B_{ab}$, is equivalent to the DBI action of an abelian gauge field with a Chern-Simons coupling $B\wedge F$. This can be regarded as the electric-magnetic duality.
We relax the Maxwell constraint $\mathcal{D}^{\alpha }\mathcal{W}_{\alpha }-\bar{\mathcal{D}}^{\dot{\alpha} }\mathcal{W}_{\dot{\alpha} }=0$ on $\mathcal{W}_{\alpha}$ in Eq. $\eqref{local B+F matter}$. We impose it by adding the following term,
 \begin{align}
S'_M=S_M +2[i \bar{\hat{\mathcal{W}}}\mathcal{W}]_F, \label{local B+F rel}
\end{align}
where $\hat{\mathcal{W}}_{\alpha }=-\frac{1}{2} \Sigma \mathcal{D}_{\alpha }\hat{V}$ and $\hat{V}$ is a real superfield. This superfield $\hat{\mathcal {W}}$ is a magnetic-dual field strength superfield of $\mathcal{W}$ because the E.O.M of $\mathcal{W}$ gives
\begin{align}
\hat{\mathcal{W}}_\alpha=\frac{i}{2}\frac{\delta S_M}{\delta \mathcal{W}^\alpha}.
\end{align}
Now, $\mathcal{W}$ in Eq. $\eqref{local B+F rel}$ should be understood as the unconstrained chiral superfield (with a spinor index). Let us see how the original system is reproduced : The variation with respect to $\hat{V}$ gives
\begin{align}
\nonumber [i \bar{\hat{\mathcal{W}}}^{\alpha }\mathcal{W}_{\alpha }]_F&=\frac{1}{2}[i\mathcal{W}_{\alpha }\Sigma (\mathcal{D}^{\alpha }\hat{V})]_F\\
\nonumber &=\frac{1}{4}[i\mathcal{W}_{\alpha }\mathcal{D}^{\alpha }\hat{V}+{\rm h.c.}]_D\\
\nonumber &=\frac{1}{4}[i(\mathcal{D}^{\alpha }\mathcal{W}_{\alpha })\hat{V}+{\rm h.c.}]_D\\
&=\frac{1}{4}[i(\mathcal{D}^{\alpha }\mathcal{W}_{\alpha }-\bar{\mathcal{D}}^{\dot{\alpha} }\mathcal{W}_{\dot{\alpha} })\hat{V}]_D , \label{Maxwell}
\end{align}
and this yields the original action $\eqref{local B+F matter}$ where ${\cal W}$ satisfies the Maxwell constraint.

Instead, if we first vary the $\mathcal{W}$, then the E.O.M of $\mathcal{W}$ reads
\begin{align}
\mathcal{W}_{\alpha }=2ig\Phi_{\alpha }-\frac{i}{2b\Lambda }\hat{\mathcal{W}}_{\alpha }. \label{til W}
\end{align}
Substituting it into Eq. $\eqref{local B+F rel}$ and redefining $\Lambda X\rightarrow X$, we obtain the dual action
\begin{align}
S'_M=2\biggl[ f(S^i)X+\frac{1}{\Lambda} \left(X+X\Sigma\left( \frac{ \bar{\omega} (\ell,S^i,\bar{S}^{\bar{j}}) \bar{X}}{\bar{\mathcal{U}}_0^2} \right) +\frac{1}{4b}\hat{\mathcal{W}}^2  \right) -2g\Phi\hat{\mathcal{W}} + \frac{M}{\Lambda ^2}X^2\biggr]_F +S_{L,M}, \label{local BF matter}
\end{align}
where we have used the following identity~\cite{Ferrara:2016een},
\begin{align}
\biggl[\Sigma \left( \frac{\omega  X\bar{X}}{\mathcal{U}_0^2\bar{\Lambda }} \right)\biggr]_F  =\biggl[ \Sigma \left( \frac{\bar{\omega} X\bar{X}}{\bar{\mathcal{U}}_0^2\Lambda} \right)\biggr]_F .
\end{align}
Note that $\bar{\Phi}\hat{\mathcal{W}}$ in Eq.~$\eqref{local BF matter}$ is invariant under the gauge transformation $\eqref{gaugetheta}$ up to the Maxwell constraint for $\hat{\mathcal{W}}$.

The Lagrange multiplier term $1/\Lambda $ in Eq.~$\eqref{local BF matter}$ gives rise to an algebraic equation, whose solution is $X=X(\hat{\mathcal{W}}, \omega, \mathcal{U}_0)$. Then, the action becomes
\begin{align}
S'_M=2[ f(S^i)X(\hat{\mathcal{W}}, \omega, \mathcal{U}_0)]_F+[-2g\Phi\hat{\cal W}]_F+S_{L,M}.
\end{align}
Note that the term proportional to $M$ in Eq.~$\eqref{local BF matter}$ automatically vanishes for the solution $X=X(\hat{\mathcal{W}}, \omega, \mathcal{U}_0)$. The first term corresponds to the DBI type action of an abelian gauge superfield $\hat{\mathcal{W}}$, which has not ${\cal F}_{ab}=F_{ab}+B_{ab}$ but $\hat{F}_{ab}$. Here $\hat{F}_{ab}$ is a field strength which belongs to $\hat{\mathcal{W}}$. The second term gives the Chern-Simons term $B\wedge \hat{F}$. This is one of the different pictures of the original action $\eqref{local B+F matter}$.  
\subsubsection*{Linear-chiral dual}
There is another dual picture of the system. We start with Eq. $\eqref{local BF matter}$ and use the linear-chiral duality~\cite{Siegel:1979ai} : The relevant part of Eq. $\eqref{local BF matter}$ is 
\begin{align}
[\mathcal{U}_0^{\frac{1}{2}}\bar{\mathcal{U}}_0^{\frac{1}{2}} \mathcal{F}(\ell, S^i,\bar{S}^{\bar{j}})]_D+2\left[\frac{1}{\Lambda}X\Sigma\left( \frac{ \bar{\omega} (\ell,S^i,\bar{S}^{\bar{j}}) \bar{X}}{\bar{\mathcal{U}}_0^2} \right) -2g\Phi\hat{\cal W}\right]_F, \label{VL}
\end{align}
For the second term, we perform the following partial integration in the same way as Eq.~$\eqref{Maxwell}$, 
\begin{align} 
\left[\frac{1}{\Lambda}X\Sigma\left( \frac{ \bar{\omega} (\ell,S^i,\bar{S}^{\bar{j}}) \bar{X}}{\bar{\mathcal{U}}_0^2} \right) -2g \Phi\hat{\cal W}\right]_F=\left[-\frac{g}{2}\hat{V}L+\frac{1}{2}\left(\frac{\bar{\omega}}{\Lambda \bar{\mathcal{U}}_0^2}+{\rm h.c.}\right)X\bar{X}\right]_D,
\end{align} 
where $L=\mathcal{D}^{\alpha }\Phi_{\alpha }+\bar{\mathcal{D}}^{\dot{\alpha }}\Phi_{\dot{\alpha} }$ and $L$ is a real linear superfield which satisfies $\Sigma(L)=\bar{\Sigma}(L)=0$.
Then one finds that with a real superfield $U$ and a chiral superfield $\varphi$ whose weights are $(2,0)$ and $(0,0)$ respectively, Eq.~$\eqref{VL}$ is equivalent to
\begin{align}
\left[\mathcal{U}_0^{\frac{1}{2}}\bar{\mathcal{U}}_0^{\frac{1}{2}}\left\{\mathcal{F}(\ell_{U}, S^i,\bar{S}^{\bar{j}})-g \hat{V}\ell_U+\left(\frac{\bar{\omega}}{\Lambda \bar{\mathcal{U}}_0^2}+{\rm h.c.}\right)|x|^2+\ell_U(\varphi +\bar{\varphi })\right\}\right]_D, 
\end{align}
where 
\begin{align}
&\ell_{U}\equiv U/\mathcal{U}_0^{\frac{1}{2}}\bar{\mathcal{U}}_0^{\frac{1}{2}} ,\\
&|x|^2\equiv \frac{X\bar{X}}{\mathcal{U}^{\frac{1}{2}}_0\bar{\mathcal{U}}_0^{\frac{1}{2}}}.
\end{align}
 The variation with respect to $\varphi $ ($\bar{\varphi }$) imposes the linearity condition on $U$, i.e.,
\begin{align}
\bar{\Sigma }U=\Sigma U=0,
\end{align}
 which leads to $U=L$ and we obtain Eq. $\eqref{VL}$. On the other hand, the E.O.M of $U$ gives 
 \begin{align}
\mathcal{F}'+\varphi +\bar{\varphi }-g \hat{V}+\left(\frac{\bar{\omega}'}{\Lambda \bar{\mathcal{U}}_0^2}+{\rm h.c.}\right)|x|^2=0,\label{EOMU}
\end{align}
where prime denotes the derivative with respect to $\ell_{U}$. In principle, we can solve this equation as
\begin{align}
\ell_{U}=&\ell_{U}(\varphi +\bar{\varphi }-g \hat{V},S^i,\bar{S}^{\bar{j}}, \mathcal{U}_0,\bar{\mathcal{U}}_0,X,\bar{X},\Lambda, \bar{\Lambda})\nonumber\\
=&\ell_1+\ell_2|x|^2,
\end{align}
where
\begin{align}
&\ell_1=\ell_U|_{x=0},\\
&\ell_2=\partial_x\partial_{\bar{x}}\ell_U|_{x=0}.
\end{align}
This is a general expression because of the nilpotency of $X$. By taking $X=0$ for both sides of Eq.~$\eqref{EOMU}$, 
\begin{align}
\mathcal{F}'|_{\ell_U=\ell_1}+(\varphi +\bar{\varphi }-g \hat{V})=0.\label{ell0}
\end{align}
We can also regard $\ell_1$ as a solution of this equation. 
By substituting the solution $\ell_U=\ell_1+\ell_2|x|^2$ into the action, we obtain
\begin{align}
\nonumber S''_M=&[\mathcal{U}_0^{\frac{1}{2}}\bar{\mathcal{U}}_0^{\frac{1}{2}}\mathcal{H}(\varphi +\bar{\varphi }-g  \hat{V},S^i,\bar{S}^{\bar{j}}, X,\bar{X},\mathcal{U}_0,\bar{\mathcal{U}}_0,\Lambda, \bar{\Lambda})]_D\\
&+2\biggl[X+\frac{1}{\Lambda} \left( f(S^i)X+\frac{1}{4b}\hat{\mathcal{W}}^2  \right) +\frac{M}{\Lambda ^2}X^2\biggr]_F . 
\end{align}
Here $\mathcal{H}$ is a real function with weights $(0,0)$, and the relation with $\mathcal{F}$ is given by
\begin{align} 
\mathcal{H}=&\left[\mathcal{F}(\ell_{U},S^i,\bar{S}^{\bar{j}})+\ell_{U}(\varphi +\bar{\varphi }-g\hat{V})+\left(\frac{\bar{\omega}}{\Lambda\bar{\mathcal{U}}_0^2}+{\rm h.c.}\right)|x|^2\right] \Biggr| _{\ell_{U}=\ell_1+\ell_2|x|^2}\nonumber\\
=&\left[\mathcal{F}(\ell_1,S^i,\bar{S}^{\bar{j}})+\ell_1(\varphi +\bar{\varphi }-g\hat{V})+\left(\frac{\bar{\omega}|_{\ell_U=\ell_1}}{\Lambda\bar{\mathcal{U}}_0^2}+{\rm h.c.}\right)|x|^2\right]\nonumber\\
&+\left[\mathcal{F}'(\ell_1,S^i,\bar{S}^{\bar{j}})+(\varphi +\bar{\varphi }-g\hat{V})\right]\ell_2|x|^2\nonumber\\
=&\left[\mathcal{F}(\ell_1,S^i,\bar{S}^{\bar{j}})+\ell_1(\varphi +\bar{\varphi }-g\hat{V})+\left(\frac{\bar{\omega}|_{\ell=\ell_1}}{\Lambda\bar{\mathcal{U}}_0^2}+{\rm h.c.}\right)|x|^2\right],
\end{align}
where we have used Eq.~$\eqref{ell0}$ in the third equality. Therefore, we finally obtain the action
\begin{align}
\nonumber S''_M&=[\mathcal{U}_0^{\frac{1}{2}}\bar{\mathcal{U}}_0^{\frac{1}{2}}H(\varphi +\bar{\varphi }-g  \hat{V},S^i,\bar{S}^{\bar{j}})]_D\\
&+2\biggl[X+\frac{1}{\Lambda} \left( f(S^i)X+X\Sigma\left( \frac{ \bar{\omega} (\ell_1,S^i,\bar{S}^{\bar{j}}) \bar{X}}{\bar{\mathcal{U}}_0^2} \right)+\frac{1}{4b}\hat{\mathcal{W}}^2  \right) +\frac{M}{\Lambda ^2}X^2\biggr]_F, \label{local FF}
\end{align}
where $H\equiv \mathcal{H}|_{X=0}$.
This is nothing but the massive DBI action derived in Refs.~\cite{Abe:2015fha,Ambrosetti:2009za,Ferrara:2015ixa}. 

The component expressions of dual actions $\eqref{local BF matter}$ and $\eqref{local FF}$ are shown in appendix~\ref{DUAL}.
\section{Discussions and summary}\label{summary}
In this paper, we have discussed the SUGRA generalization of SUSY DBI action coupled to a two-form field. Based on conformal SUGRA technique, we have considered possible matter coupled extension and shown that it can be realized in any off-shell formulations with different auxiliary fields. This fact is contrast to our previous work~\cite{Aoki:2016cnw}, where we found the DBI action of a complex scalar field cannot be realized in old minimal formulation as a naive SUGRA embedding of SUSY models in Ref.~\cite{Rocek:1997hi}.\footnote{The action is also a DBI action with a linear superfield as our present model. However, these are essentially different from each other since the present one is realized with prepotential but the previous one with a (gauge invariant) linear superfield. The former is dual to a massive vector superfield, and the latter is dual to a chiral superfield.} In addition to this, there exists another model, which is realized only in new minimal formulation~\cite{Farakos:2012je}. It would be interesting to explore the (physical) reasons behind that if any, in the study of SUSY higher-order derivative models. Also in such models, it might be important to clarify what formulations of SUGRA are allowed, as we have done in this work, since it is expected that duality transformation between different formulations~\cite{Ferrara:1983dh} does not work in the existence of higher-order derivative couplings.

We have also discussed the duality relations of DBI actions in terms of superfields. We have shown that DBI action of generalized gauge invariant two-form ${\mathcal {F}_{ab}}=F_{ab}+B_{ab}$ is equivalent to the DBI action of $F_{ab}$ with a SUGRA extension of $B\wedge F$, or to the massive DBI action discussed in Refs.~\cite{Abe:2015fha,Ambrosetti:2009za,Ferrara:2015ixa}. It is remarkable that we find these duality relations, which are expected from the SUSY case~\cite{Kuzenko:2000uh,Ambrosetti:2009za,Ferrara:2015ixa}, hold independently of the choice of the compensator superfields. 

As we have discussed in Sec.~\ref{intro}, although the D-brane action coupled to SUGRA has been shown and our result would also contribute to such an attempt, the SUGRA completion of the Dp-brane requires further investigation. Recently, it has been shown that the anti-D3-brane under a specific condition would be described as the Volkov-Akulov(VA) nilpotent superfield~\cite{Kallosh:2014wsa,Bergshoeff:2015jxa,Bandos:2015xnf}. In Refs.~\cite{Vercnocke:2016fbt,Kallosh:2016aep}, the relation between the DBI-VA action~\cite{Bergshoeff:2013pia} and constrained superfields has been also investigated, which would clarify the SUGRA realization of D-brane actions. 

Another important thing is the cosmological applications of DBI(-VA) action. One of the directions is the massive vector inflation, where the Stuckelberg superfield plays the role of the inflaton~\cite{Ferrara:2013rsa}. As realized in Ref.~\cite{Abe:2015fha}, the inclusion of DBI corrections to the potential can make the inflaton potential flatter than that without higher order corrections. The model in this paper is dual to the massive DBI model and must be applicable to such model construction. The other direction is the use of DBI action as the stabilizer/Goldstino superfield, which is described by constrained superfields~\cite{Antoniadis:2014oya,Ferrara:2014kva,Kallosh:2014via}. In Ref.~\cite{Kallosh:2014wsa}, the D3-brane action with orientifold conditions gives the VA action. It would be rather important to clarify the relation between the nilpotent superfield and our DBI action in terms of superfields. We expect that such a relation would be found since the VA action has SUSY nonlinearly. In our construction of the DBI action shown in Sec.~\ref{extension}, we have shown that the DBI constraint has the hidden nilpotency $X^2=0$, which would be a key to understand the relation. We will investigate these aspects of DBI action in SUGRA elsewhere.
\section*{Acknowledgment}
SA thanks Hiroyuki Abe and Ryo Yokokura for useful discussions. YY would like to thank Renata Kallosh for useful comments. The work of SA is supported by Research Fellowships of Japan Society for the Promotion of Science for Young Scientists Grant Numbers 16J06569. The work of YY is supported by SITP and the NSF grant PHY-1316699.
\begin{appendix}
\section{Integration of auxiliary fields and superconformal gague fixing}\label{AUX}
Here, we show some details for deriving the component action $\eqref{final}$ from $\eqref{local B+F matter}$ in old minimal case with 
\begin{align}
\mathcal{U}_0=S_0\bar{S}_0.\label{choiceu}
\end{align}

For notational simplicity,  we divide our action $\eqref{local B+F matter}$ into two parts
\begin{align}
S_M=S_{L,M}+S_{DBI,M},  \label{local B+F matterT} 
\end{align}
where $S_{L,M}$ denotes Eq. $\eqref{SLM}$ with a possible superpotential $\eqref{W}$ of matter multiplets $S^i$, i.e.,
\begin{align}
S_{L,M}=[S_0\bar{S}_0\mathcal{F}(\ell, S^i,\bar{S}^{\bar{j}})]_D +2[S_0^3W(S^i)]_F, \ \ \ \ {\rm{with}}\ \  \ell=\frac{L}{S_0\bar{S}_0}. \label{SLMW}
\end{align}
Then, we first discuss the DBI part $\mathcal{L}_{DBI,M}$, and derive the on-shell action. Note that procedures obtaining on-shell actions, that is, eliminating auxiliary fields, can be done separately as long as matter chiral multiplets $S^i$ are gauge singlets, which we assume here for simplicity. Even if there exist charged matters, the procedure is almost the same (see ,e.g.,~Refs.~\cite{Abe:2015nxa,Abe:2015fha}).

Applying the superconformal action formulas $\eqref{Fformula}$, $\eqref{Dformula}$ and multiplication rule $\eqref{formulaF}$ to $S_{DBI,M}$, we obtain the bosonic components,
\begin{align}
\nonumber \mathcal{L}_{DBI,M}=&( f+\Lambda  ) F_X +\frac{\omega |F_X|^2}{|S_0|^4}(\Lambda+\bar{\Lambda}) +\\
&b\Lambda \biggl[ \frac{1}{2}\mathcal{F}_{ab}\mathcal{F}^{ab}-\frac{1}{2}\mathcal{F}_{ab}\tilde{\mathcal{F}}^{ab}+\left(  iD-\frac{g}{2}(C+iE)\right)^2\biggr] +{\rm{h.c.}}. \label{local B+F Mcom1}
\end{align}
In this action, we need to eliminate auxiliary fields, $F_X$, $\Lambda $ and $D$. 
One can integrate out $F_X$ and obtain the following Lagrangian, with $\Lambda =\Lambda _R+i\Lambda _I $, 
\begin{align}
\nonumber \mathcal{L}_{DBI,M}=&-\frac{|S_0|^4}{4\omega }\frac{(f_R+\Lambda _R)^2+(f_I+\Lambda _I)^2}{\Lambda _R}+b\left( \mathcal{F}_{ab}\mathcal{F}^{ab}+\frac{1}{2}(gC)^2-2\left(  D-\frac{g}{2}E\right)^2 \right) \Lambda _R\\
&-b\left(  i\mathcal{F}_{ab}\tilde{\mathcal{F}}^{ab}-2gC\left( D-\frac{g}{2}E \right)\right) \Lambda _I .\label{local B+F Mcom2}
\end{align}
Next, we algebraically solve the E.O.Ms for $\Lambda _R$ and $\Lambda _I$, which yield
\begin{align}
\nonumber  \Lambda _R=&\pm f_R \biggl[ 1-\frac{4b\omega }{|S_0|^4}\mathcal{F}_{ab}\mathcal{F}^{ab}-\frac{2b\omega }{|S_0|^4}\left((gC)^2- 4\left(  D-\frac{g}{2}E\right)^2\right) \\
&-\frac{4(b\omega )^2}{|S_0|^8}\left(  i\mathcal{F}_{ab}\tilde{\mathcal{F}}^{ab}- 2gC\left(  D-\frac{g}{2}E\right)\right)^2 \biggr] ^{-\frac{1}{2}}, \\
\Lambda _I=&-f_I-\frac{2b\omega }{|S_0|^4}\left(  i\mathcal{F}_{ab}\tilde{\mathcal{F}}^{ab}- 2gC\left(  D-\frac{g}{2}E\right) \right) \Lambda _R.
\end{align} 
Substituting them into Eq. $\eqref{local B+F Mcom2}$, we obtain
\begin{align}
\nonumber \mathcal{L}_{DBI,M}=&-\frac{|S_0|^4f_R}{2\omega }+ibf_I\mathcal{F}_{ab}\tilde{\mathcal{F}}^{ab}-2f_IbgC\left(  D-\frac{g}{2}E\right)\\
\nonumber &\pm \frac{|S_0|^4f_R}{2\omega }\biggl[ 1-\frac{4b\omega }{|S_0|^4}\mathcal{F}_{ab}\mathcal{F}^{ab}-\frac{2b\omega }{|S_0|^4}\left((gC)^2- 4\left(  D-\frac{g}{2}E\right)^2\right) \\
&-\frac{4(b\omega )^2}{|S_0|^8}\left(  i\mathcal{F}_{ab}\tilde{\mathcal{F}}^{ab}- 2gC\left(  D-\frac{g}{2}E\right)\right)^2 \biggr] ^{\frac{1}{2}} . \label{local B+F Mcom3}
\end{align}
Finally, we eliminate $D$ whose E.O.M is given by
\begin{align}
 \nonumber  D-\frac{g}{2}E=&-\frac{i\frac{b\omega}{|S_0|^4} gC}{1-\frac{2b\omega }{|S_0|^4}(gC)^2}\mathcal{F}_{ab}\tilde{\mathcal{F}}^{ab} \\
 &\pm \frac{f_I}{2}gC\sqrt{\frac{1}{f_R^2-\frac{2b\omega }{|S_0|^4}(gC)^2(f_R^2+f_I^2)}}\times \sqrt{-{\rm{det}}\left\{ g_{ab}+\sqrt{\frac{-8b\omega /|S_0|^4}{1-\frac{2b\omega }{|S_0|^4}(gC)^2}}\mathcal{F}_{ab}\right\} }.
\end{align}
Then, the on-shell superconformal action is
\begin{align}
\nonumber \mathcal{L}_{DBI,M}=&-\frac{|S_0|^4f_R}{2\omega }+\frac{2ibf_I}{1-\frac{2b\omega }{|S_0|^4}(gC)^2}\mathcal{F}_{ab}\tilde{\mathcal{F}}^{ab} \\
&+ \frac{|S_0|^4}{2\omega }\sqrt{f_R^2-\frac{2b\omega }{|S_0|^4}(gC)^2(f_R^2+f_I^2)}\times \sqrt{-{\rm{det}}\left\{ g_{ab}+\sqrt{\frac{-8b\omega /|S_0|^4}{1-\frac{2b\omega }{|S_0|^4}(gC)^2}}\mathcal{F}_{ab}\right\} }. \label{local B+F Mcom4}
\end{align}

Let us move to the part $\mathcal{L}_{L,M}$. This part was studied well in Ref.~\cite{Derendinger:1994gx}, so we only comment on superconformal gauge fixing and give the final on-shell Lagrangian here. From D-term formula $\eqref{Dformula}$, the coefficient of Ricci scalar is found to be
\begin{align}
\frac{1}{2}R\left(  -\frac{2}{3}|S_0|^2(\mathcal{F}-C\mathcal{F}_C)\right) ,
\end{align}
where $\mathcal{F}_C=\partial \mathcal{F}/\partial C$. The second term including $\mathcal{F}_C$ comes from the fact that $\Box C$ in Eq.~$\eqref{embedl}$ contains the Ricci scalar. Then, we impose superconformal gauge fixing conditions as
\begin{align}
-\frac{2}{3}|S_0|^2(\mathcal{F}-C\mathcal{F}_C) &=1, \label{Dcondition}\\
S_0&=\bar{S}_0,\label{Acondition}\\
b_{\mu}&=0.\label{Kcondition}
\end{align}
Here, $b_{\mu}$ is the gauge field of dilatation symmetry. The first condition eliminates dilatation symmetry and ensures the action in the Einstein frame. The second and third conditions correspond to fixings of $U(1)_A$ symmetry and conformal boost, respectively.
One can see that Eq. $\eqref{Dcondition}$ is an implicit equation with respect to $S_0$ since $\mathcal{F}$ has dependence on it through $\ell$, then we do not have an explicit solution in general. This fact may result in complicated component form of $\mathcal{L}_{L,M}$. We use the specific choice of $\mathcal{F}$ as Eq. $\eqref{ellG}$ following Ref.~\cite{Derendinger:1994gx}, though Eq. $\eqref{Dcondition}$ is still implicit for $S_0$. With this choice, however, much simplification occurs after integrating out auxiliary fields $F_0, F^i, \mathcal{A}_{\mu}$, which are $F$ components of $S_0$ and $S^i$, a gauge field for $U(1)_A$ symmetry, respectively. Then we finally obtain the expression~\cite{Derendinger:1994gx},
\begin{align}
\mathcal{L}_{L,M}=\frac{1}{2}R -K_{i\bar{j}}\partial _{\mu}S^i \partial ^{\mu}\bar{S}^{\bar{j}} -\frac{3}{4C^2}\frac{2G'+yG''}{G'+yG''}\left( \partial_{\mu}C \partial^{\mu}C -B_{\mu}B^{\mu}\right) -V_F, \label{SLMcomp}
\end{align} 
where the F-term scalar potential $V_F$ is given by Eq. $\eqref{V}$.

The sum of Eq. $\eqref{local B+F Mcom4}$ and Eq. $\eqref{SLMcomp}$ leads to the action $\eqref{final}$.


\section{Component expressions of dual actions}\label{DUAL}
For completeness, we show components of dual actions $\eqref{local BF matter}$ and $\eqref{local FF}$.  We also follow the choice $\eqref{choiceu}$ here.

We start from Eq. $\eqref{local BF matter}$. The part $\mathcal{S}_{L,M}$ is same as Eq. $\eqref{SLMW}$ with components $\eqref{SLMcomp}$, then we discuss the remaining part $\mathcal{S}_{DBI,M}$. 
Under following field redefinitions, $fX= \tilde{X}$, $\frac{1}{\Lambda }= \tilde{\Lambda} $, $\frac{M\tilde{\Lambda} ^2}{ f(S^i) ^2}=\tilde{M}$ and dropping tildes again, the DBI part of Eq. $\eqref{local BF matter}$ is equivalent to 
\begin{align}
S_{DBI,M}=2\biggl[\frac{1}{ f(S^i)}X+\Lambda  \left( X+X\Sigma\left( \frac{ \bar{\omega} (\ell,S^i,\bar{S}^{\bar{j}}) \bar{X}}{| f(S^i)|^2\bar{\mathcal{U}}^2_0} \right) +\frac{1}{4b}\hat{\mathcal{W}}^2  \right) -2g\bar{\Phi}\hat{\mathcal{W}} + MX^2\biggr]_F . \label{local BF matter2}
\end{align}
Then, one can regard the first term and parts which are proportional to $\Lambda$ as Eq. $\eqref{local B+F matter}$ with the following  replacements,
\begin{align}
f\rightarrow \frac{1}{f}, \ \ \ \ \omega \rightarrow \frac{\bar{\omega }}{|f|^2},\ \ \ \ b\rightarrow \frac{1}{4b}, \ \ \ \ \mathcal{W}+\Phi \rightarrow \hat{\mathcal{W}},
\end{align} 
thus the things to do are almost the same. The components are given by
\begin{align}
\nonumber \mathcal{L}_{DBI,M}=&\frac{i}{2}g\hat{F}_{ab}\tilde{B}^{ab}-\frac{i}{4b}\frac{f_I}{f_R^2+f_I^2}\hat{F}_{ab}\tilde{\hat{F}}^{ab}\\
&-\frac{|S_0|^4f_R}{2\omega }\biggl[ 1- \sqrt{1-\frac{b\omega }{2|S_0|^4}\frac{f_R^2+f_I^2}{f_R^2}(gC)^2} \sqrt{-{\rm{det}}\left\{ g_{ab}+\sqrt{\frac{-2\omega }{b|S_0|^4(f_R^2+f_I^2)}}\hat{F}_{ab}\right\} } \biggr] , \label{local BF Mcom4}
\end{align}
We have also assumed that $\omega $ is real here. Then, the total on-shell Lagrangian of Eq. $\eqref{local BF matter}$ is given by Eq. $\eqref{SLMcomp}$+Eq. $\eqref{local BF Mcom4}$.

Finally, we discuss the action $\eqref{local FF}$.
In this case, there is no linear multiplet since it has been already converted to a chiral multiplet $\varphi $, thus we can perform the well-known superconformal gauge fixings, in contrast to the above discussions. As we mentioned before, this action describes massive vector field with DBI corrections after imposing gauge fixing condition $\varphi =0$. The massive vector multiplet $\hat{V}$ has following bosonic components,
\begin{align}
\hat{V}=\{\hat{C}, ..., \hat{H}, \hat{K}, \hat{B}_a, ..., \hat{D}  \},
\end{align}
where all of them are real and ellipses denote fermion parts. The physical bosonic fields are $\hat{C}$ and $ \hat{B}_a$, and others are auxiliary fields. It is not difficult to eliminate all of the auxiliary fields including $F_X,\Lambda,F_0,F^i,\mathcal{A}_{\mu}$\footnote{The matter coupled Lagrangian without DBI corrections can be found in Ref.~\cite{Aldabergenov:2016dcu}.}. Then, we fix the dilatation symmetry by
\begin{align}
-\frac{2}{3}|S_0|^2H=1.
\end{align}
One can see that this condition can be solved with respect to $S_0$. The conditions for $U(1)_A$ symmetry and conformal boost are the same as Eqs. $\eqref{Acondition}$ and $\eqref{Kcondition}$. After that, we obtain the on-shell Lagrangian,
\begin{align}
\nonumber \mathcal{L}=&\frac{1}{2}R-\mathcal{J}_{i\bar{j}}\partial _{\mu}S^i \partial ^{\mu} \bar{S}^{\bar{j}}-\frac{1}{4}\mathcal{J}_{\hat{C}\hat{C}}\partial _{\mu}\hat{C} \partial ^{\mu} \hat{C} -\frac{1}{4}\mathcal{J}_{\hat{C}\hat{C}}\hat{B}_{\mu}\hat{B}^{\mu} \\
\nonumber &+\left\{  -\frac{1}{2}\mathcal{J}_{\hat{C}i} \partial _{\mu}\hat{C} \partial ^{\mu}S^i -\frac{i}{2}\mathcal{J}_{\hat{C}i}\hat{B}_{\mu}\partial ^{\mu}S^i +{\rm{h.c.}}\right\}-V_F\\
\nonumber &-\frac{i}{4b}\frac{f_I}{f_R^2+f_I^2}\hat{F}_{ab}\tilde{\hat{F}}^{ab}-\frac{f_R e^{\frac{2}{3}\mathcal{J}}}{2\omega }\\
& + \frac{f_R e^{\frac{2}{3}\mathcal{J}}}{2\omega } \sqrt{1-\frac{b\omega }{2}\frac{f_R^2+f_I^2}{f_R^2}\mathcal{J}_{\hat{C}}^2e^{-\frac{2}{3}\mathcal{J}}} \times \sqrt{-{\rm{det}}\left\{ g_{ab}+\sqrt{\frac{-2\omega }{b(f_R^2+f_I^2)}e^{-\frac{2}{3}\mathcal{J}}}\hat{F}_{ab}\right\} } 
\end{align}
where we introduced a real function $\mathcal{J}$ by
\begin{align}
\mathcal{J}=-3\log \left(  -\frac{2}{3}H \right) .
\end{align}
The subscripts of $\mathcal{J}$ denote derivatives with respect to $\hat{C}, S^i, \bar{S}^{\bar{j}}$, in the same way as $K$ in Eq. $\eqref{final}$ .   
The F-term scalar potential $V_F$ is given by
\begin{align}
\nonumber V_F=&e^{\mathcal{J}}\Biggl[\left( \mathcal{J}_{i\bar{j}}-\frac{\mathcal{J}_{\hat{C}i}\mathcal{J}_{\hat{C}\bar{j}}}{\mathcal{J}_{\hat{C}\hat{C}}} \right)^{-1}\left( W_i+ \mathcal{J}_iW -\frac{\mathcal{J}_{\hat{C}}\mathcal{J}_{\hat{C}i}}{\mathcal{J}_{\hat{C}\hat{C}}}W\right) \left( \bar{W}_{\bar{j}}+ \mathcal{J}_{\bar{j}}\bar{W} -\frac{\mathcal{J}_{\hat{C}}\mathcal{J}_{\hat{C}\bar{j}}}{\mathcal{J}_{\hat{C}\hat{C}}}\bar{W}\right) \\
&-3|W|^2+\frac{\mathcal{J}_{\hat{C}}^2}{\mathcal{J}_{\hat{C}\hat{C}}}|W|^2\Biggr] .
\end{align}
This action reduces to the matter coupled massive vector action obtained in Ref.~\cite{Aldabergenov:2016dcu} in the limit $\omega \rightarrow 0$ and $f \rightarrow 1$, where DBI corrections disappear. 
\end{appendix}

\end{document}